\documentclass[a4paper,11pt]{article}

\usepackage[T1]{fontenc}         % Codifca caratteri
\usepackage[utf8]{inputenc}      % Imposta la codifica di input per interpretare i caratteri immessi nell'editor; opzioni: utf8, utf8x, latin1, ansinew
\usepackage{microtype}           % Rende la distribuzione delle lettere più omogenea
\usepackage{lmodern}             % Caricamento del font principale
\usepackage{textcomp}            % Simboli aggiuntivi molto comuni
\usepackage[frenchb,ngerman,italian,english]{babel}  % Opzioni lingue; NB:l'ultima lingua risulta quella predefinita

\usepackage{amscd,amsmath,amsfonts,amssymb,amsthm}     % Pacchetti per la matematica
\usepackage{bm}                  % Caratteri matematici in grassetto (caratteri romani e greci)
\usepackage{braket}              % Per le parentesi con spigolo e per definire insiemi numerici (comunque parentesi in generale)
\usepackage[mathcal]{euscript}   % Per utilizzare il font \mathcal{}
\usepackage{mathrsfs}            % Per utilizzare il font matematico ``manoscritto'' (\mathscr)
\usepackage{mathtools}           % IMPORTANTE! Serve per generare nuovi comandi matematici
\usepackage{tensor}              % Per scrivere correttamente apici e pedici dei tensori: M\indices{^a_b^{cd}_e} oppure \tensor[^a_b^c_d]{M}{^a_b^c_d} oppure \tensor*{M}{*^{i_1}_{m_1}^{i_2}_{m_2}^{i_3}_{m_3}^{i_4}_{m_4}}
\usepackage[output-decimal-marker={.}]{siunitx} % Unità di misura SI con possibilità di cambiare la virgola col punto e viceversa; esempio: \SI{23,4}{kg.m.s^{-2}} , \SI{0,8 e-15}{m} , \si{J.mol^{-1}.K^{-1}} ovvero solo per l'unità e non il numero
\usepackage[only,llbracket,rrbracket]{stmaryrd} % Fonts aggiuntivi per creare il comando \jump (operatore ``salto'')
\usepackage{upgreek}             % Caratteri greci ``upright''

\usepackage{booktabs}            % Migliora la composizione delle tabelle
\usepackage{caption}
\usepackage{graphicx}            % Supporto per le immagini (.pdf, .png, .jpg, .eps); ES: \usepackage[pdftex]{graphicx} ; NB: AUTOMATICAMENTE CARICATO CON TIKZ
\usepackage{multirow}            % Supporto per celle multiriga
\usepackage{rotating}            % Per ruotare le tabelle e le figure, crea gli ambienti \begin{sidewaystable} e \begin{sidewaysfigure}
\usepackage{subfig}              % Sottofigure, sottotabelle
\usepackage{tabularx}            % Tabelle di larghezza impostata dall'utente
\usepackage[svgnames,dvipsnames,table]{xcolor}  % Gestisce i colori; senza opzioni si hanno a disposizione 19 colori nominativi
\graphicspath{{./}{./figures/}}  % Percorso predefinito per le figure

\usepackage[affil-it,auth-sc]{authblk}           % Per definire gli istituti di appartenenza degli autori
\usepackage[top=25mm,bottom=25mm,left=25mm,right=25mm,heightrounded]{geometry}    % Per modificare la geometria della pagina
\usepackage[font=small,noorphans]{quoting}       % Citazioni di parti di testo: \begin{quoting} [...] \end{quoting}

\begin{document}
\selectlanguage{english}

\title{Plastically-driven variation of elastic stiffness in green bodies during powder compaction. \\ Part II: Micromechanical modelling}

\author[1]{L.P. Argani}
\author[1]{D. Misseroni}
\author[1]{A. Piccolroaz}
\author[2]{D. Capuani}
\author[1]{D. Bigoni\footnote{Corresponding author. Phone:\,+39\,0461\,282507; E-mail:\,bigoni@ing.unitn.it; Fax:\,+39\,0461\,282599.}}

\affil[1]{DICAM, University of Trento, via~Mesiano~77, I-38123 Trento, Italy.}
\affil[2]{Department of Architecture, University of Ferrara, Via Quartieri 8, 44100 Ferrara, Italy}

\date{}
\maketitle

\begin{abstract}
\noindent
A micromechanical approach is set-up to analyse the increase in elastic stiffness related to development of plastic deformation (the elastoplastic coupling concept) occurring during the compaction of a ceramic powder.
Numerical simulations on cubic (square for 2D) and hexagonal packings of elastoplastic cylinders and spheres validate both the variation of the elastic modulus with the forming pressure and the linear dependence of it on the relative density as experimentally found in Part~I of this study, while the dependence of the Poisson's ratio on the green's density is only qualitatively explained.
\end{abstract}


%======================================================================================================================
% SEZIONE 1 : INTRODUZIONE
%======================================================================================================================
\section[Introduction]{Introduction}
  \label{Sezione-01}

\begin{figure}[!htb]
\renewcommand{\figurename}{\footnotesize{Fig.}}
\begin{center}
 \includegraphics[width=12 cm]{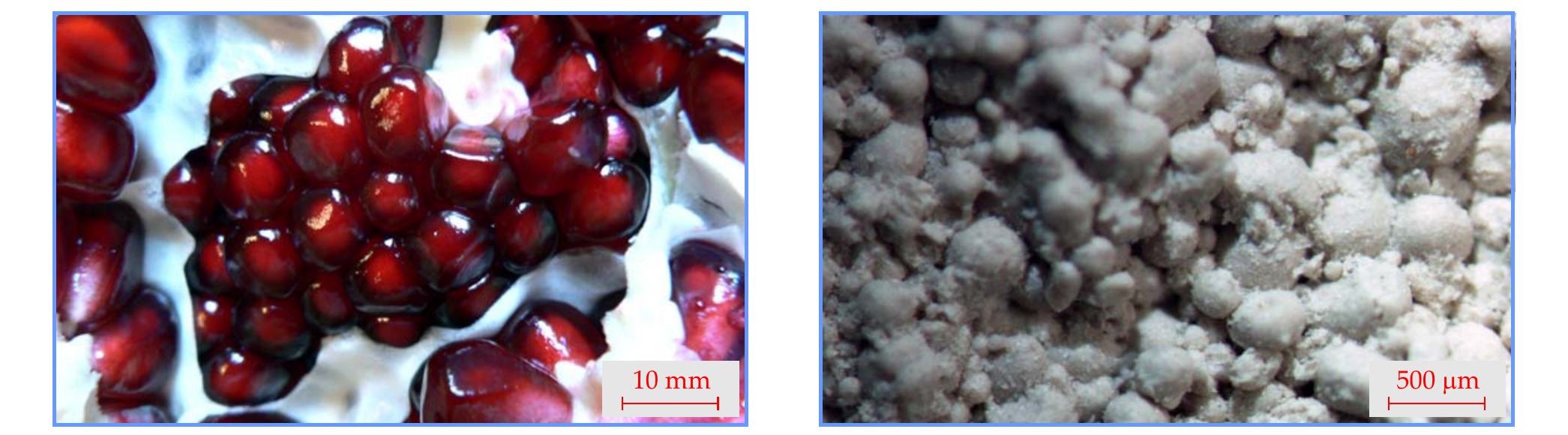}
\caption{\footnotesize Examples of packaged spherical particles in nature (pomegranate seeds, left, photo taken with a Panasonic DMC-FZ5 digital camera) and in industry (an aluminum silicate spray dried powder, right, photo kaken with 
a Nikon SMZ-800 optical microscope equipped with DSF1i camera head
)}
\label{melagrana}
\end{center}
\end{figure}

Densification of metal as well as ceramic powders is a problem connected with a strong industrial interest, so that the micromechanics of this process has been the focus of a number of investigations (almost all addressed to metal particles, while the akin problem of ceramic powders has been much less investigated).
Grains have been usually assumed as spherical (or cylindrical for simplicity), so that micromechanics explains how plasticity and increase of contact areas between particles influence the overall stress/strain behaviour.
The analysis of this problem sheds light on the macroscopic constitutive modelling of the powder, to be employed in the design of moulds to form  green pieces with desired shape.
The compaction problem is also of great academic interest in several fields, including biomechanics, where it traces back to the famous \lq Histoire Naturelle' by the Count de Buffon, who reports on a (probably \lq thought') experiment with peas: 
\begin{quoting}
\textit{Qu'on remplisse un vaisseau de pois, ou plut\^{o}t de quelqu'autre graine cylindrique, et qu'on le ferme exactement apr\`es y avoir vers\'e autant d'eau [\dots\negthinspace]; qu'on fasse boullir cette eau, tous ces cylindres deviendront des colonnes \`a six pans. On en voit clairement la raison, qui est purament m\'ecanique; chaque graine, dont la figure est cylindrique, tend par son reflement \`a occuper le plus d'espace possible dans un espace donn\'e, elles deviennent donc toutes n\'ecessairement hexagones par la compression r\'eciproque.}
\end{quoting}
This is an example of compaction of a package of spheres (figure \ref{melagrana}), later continued by D'Arcy Thompson in his \textit{On Growth and Form} and others.

%%%%%%%%%% FIG %%%%%%%%%%
%%%%%%%%%% END FIG %%%%%%%%%%

Micromechanical models of powder compaction have been developed so far for a cubic (square in 2D) geometry of spheres~\cite{Ogbonna-Fleck:1995, Mesarovic-Fleck:2000, Larsson-Olsson:2015, Harthong-Jerier-Doremus-Imbault-Donze:2009} or cylinders~\cite{Redanz-Fleck:2001, Sridhar-Fleck-Akisanya:2001, Akisanya-Cocks-Fleck:1994, Akisanya-Cocks:1995} in frictionless contact, and friction between grains has also been considered for the latter geometry~\cite{Redanz-Fleck:2001}.
Random packing of cylinders and spheres have been analyzed respectively in~\cite{Redanz-Fleck:2001} and~\cite{Harthong-Jerier-Doremus-Imbault-Donze:2009}.
All the above-mentioned investigations, in which the spheres and the cylinders are modelled within the framework of the J$_2$-flow theory of plasticity with linear hardening or perfectly plastic behaviour (figure~\ref{fig:legame-elastoplastico}), are all focused on the determination of the yield surface at different stages of compaction.

The objective of the present article is to investigate how the \textit{plastic deformation of grains} during compaction influences the macroscopic \textit{elastic} response of the material, an aspect never considered before, but central in the development of elastoplastic coupling (see Part~I of this study).
To this purpose, 2D (plane strain) and 3D square/cubic and hexagonal packings of cylindrical and spherical grains are considered (figure~\ref{fig:impaccamento}).
Although detailed information on the constitutive law valid for the grains is not available, these are modelled via von Mises perfect or linear hardening plasticity, which is typical of a basic and simple mechanical behaviour.
Representative volume elements of the cylinder and sphere packings are deformed to model the state of uniaxial strain achieved in a cylindrical rigid die and the mean stress/mean strain behaviour is numerically determined using the finite element program Abaqus Unified FEA$^\circledR$.
Once the uniaxial strain compaction has been completed, the representative element is unloaded and reloaded under uniaxial stress to evaluate the average Young modulus and Poisson's ratio of the material.
In this way it is possible to determine the variation of the elastic modulus with the forming pressure and the dependence of the elastic modulus on the density.
These evaluations validate the experimental results presented in the Part~I of this study.
The micromechanical evaluation of the Poisson's ratio is more complicated than that of the elastic modulus.
In this case, the results from micromechanics correctly explain the qualitative increase of the Poisson's ratio with the forming pressure, but the values are not tight to experimental results.

The dependence of elastic stiffness on the level of plastic deformation is a crucial aspect of elastoplastic modelling of geological and granular materials, including ceramic, metal powders, and greens.
Results provided in the present article explain the plastic micromechanisms inducing elastic stiffening during compaction of ceramic powders.

%%%%%%%%%%--FIGURA--%%%%%%%%%%--FIGURA--%%%%%%%%%%%%%%%%%%%%%%%%%%%%%%%%%%%%%%%
%{\tikzset{external/figure name={Meccanismo-Resistenza_}} % Nome per le figure esternalizzate
\begin{figure}[tp]
\centering
\subfloat[\emph{Unit cell undeformed configuration.}]{\label{fig:Meccanismo-Resistenza-a}
\includegraphics[width=0.225\columnwidth,keepaspectratio]{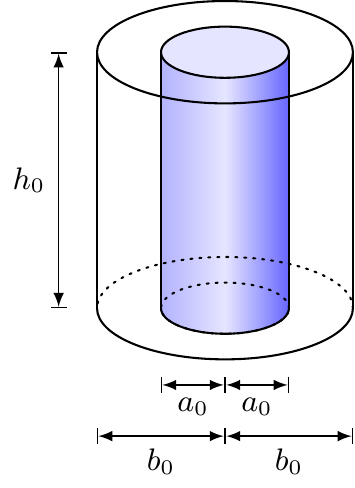}
}
\qquad
\subfloat[\emph{Unit cell deformed configuration.}]{\label{fig:Meccanismo-Resistenza-b}
\includegraphics[width=0.2156\columnwidth,keepaspectratio]{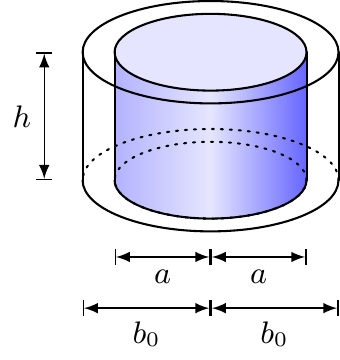}
}
\caption[A toy model to explain elastic stiffening due to plastic deformation]{A toy model to explain elastic stiffening due to plastic deformation.
The elastic circular cylinder of initial radius $ a_{0} $, height $ h_{0} $, and elastic modulus $ E $ is coaxial to the unit cell of radius $ b_{0} $.
Upon axial plastic deformation, the inner cylinder has a radius $ a $ and height $ h $ ($ a > a_{0} $ and $ h < h_{0} $).
If the plastic strain is isochoric $ a^{2} h = a_{0}^{2} h_{0} $, so that the new geometry will result elastically stiffer than the initial one.}
\label{modellino}
\end{figure}
%} % Chiusura assegnazione nomi
%%%%%%%%%%--FINE-FIGURA--%%%%%%%%%%--FINE-FIGURA--%%%%%%%%%%%%%%%%%%%%%%%%%%%%%

%======================================================================================================================
% SEZIONE 2 : ACCOPPIAMENTO ELASTOPLASTICO
%======================================================================================================================
\section[A toy mechanical model to explain elastoplastic coupling]{A toy mechanical model to explain elastoplastic coupling}
  \label{Sezione-02}

Before to set up the micromechanical model for the qualitative and quantitative explanation of elastoplastic coupling, a simple mechanical model is presented with the aim of providing a simple explanation of the phenomenon.
The model is intended only to shed light on the mechanism of increase in elastic stiffness due to plastic deformation and not to provide a quantitative evaluation.

Referring to an elastoplastic circular cylinder of initial height $ h_{0} $ and cross section of radius $ a_{0} $, this is inserted in a larger and coaxial cylindrical unit cell with cross section of radius $ b_{0} > a_{0} $, so that when the cylinder is subject to a force  $ F $ (positive when tensile), the nominal stress is
$ \sigma_{\textup{n}} = F/(\pi b_{0}^{2}) $, while the effective stress is $ \sigma_{\textup{e}} = F/(\pi a_{0}^{2}) $.
Assuming that incompressible axial plastic deformation $ \varepsilon_{\textup{p}} $ has brought the cylinder to a new height $ h $ and radius $ a $, isochoricity implies $ a^{2} =  h_{0} a_{0}^{2}/h = a_{0}^{2}/(1+\varepsilon_{\textup{p}}) $.
The axial plastic deformation $ \varepsilon_{\textup{p}} $ can be expressed in terms of void ratio as
  \begin{equation}
    \label{epsilon}
      \varepsilon_{\textup{p}} = \frac{ e-e_{0} }{ 1+e_{0} } \, ,
  \end{equation}
where $ e_{0} = (b_{0}^{2} - a_{0}^{2})/a_{0}^{2} $ is the initial void ratio and $ e = (b_{0}^{2} - a^{2})/a^{2} $ is the current void ratio.

If the deformed cylinder is now loaded with a force $F$, the nominal stress remains equal to $ \sigma_{\textup{n}} $ (because the radius of the unit cell does not change), but the deformation is $ \varepsilon_{\textup{c}} = F/(E\pi a^{2}) $, so that the apparent elastic modulus defined as $ \bar{E} = \sigma_{\textup{n}}/\varepsilon_{\textup{c}} $ is
  \begin{equation}
    \label{cazzata}
      \bar{E}(\varepsilon_{\textup{p}}) = E \frac{ a^{2} }{ b_{0}^{2} }
              = E \frac{ a_{0}^{2} }{ b_{0}^{2} (1+\varepsilon_{\textup{p}}) } \, .
  \end{equation}
Equation~\eqref{cazzata} is not expected to realistically represent the variation in elastic stiffness of a ceramic powder, but provides a simple model to understand the elastoplastic coupling effect at the microscale.
In fact, for a compressive (and therefore negative) plastic deformation $ \varepsilon_{\textup{p}} $, the apparent elastic modulus of the material $ \bar{E} $ increases, as it happens in a ceramic or metallic powder.

%%%%%%%%%%--FIGURA--%%%%%%%%%%--FIGURA--%%%%%%%%%%%%%%%%%%%%%%%%%%%%%%%%%%%%%%%
%{\tikzset{external/figure name={Tipologia-impaccamento-2D_}} % Nome per le figure esternalizzate
\begin{figure}[tp]
\centering
\subfloat[\emph{2D-Square packing, with the unit cell and the primitive cell shown red.}]{\label{fig:impaccamento-a}
\includegraphics[width=0.40\columnwidth,keepaspectratio]{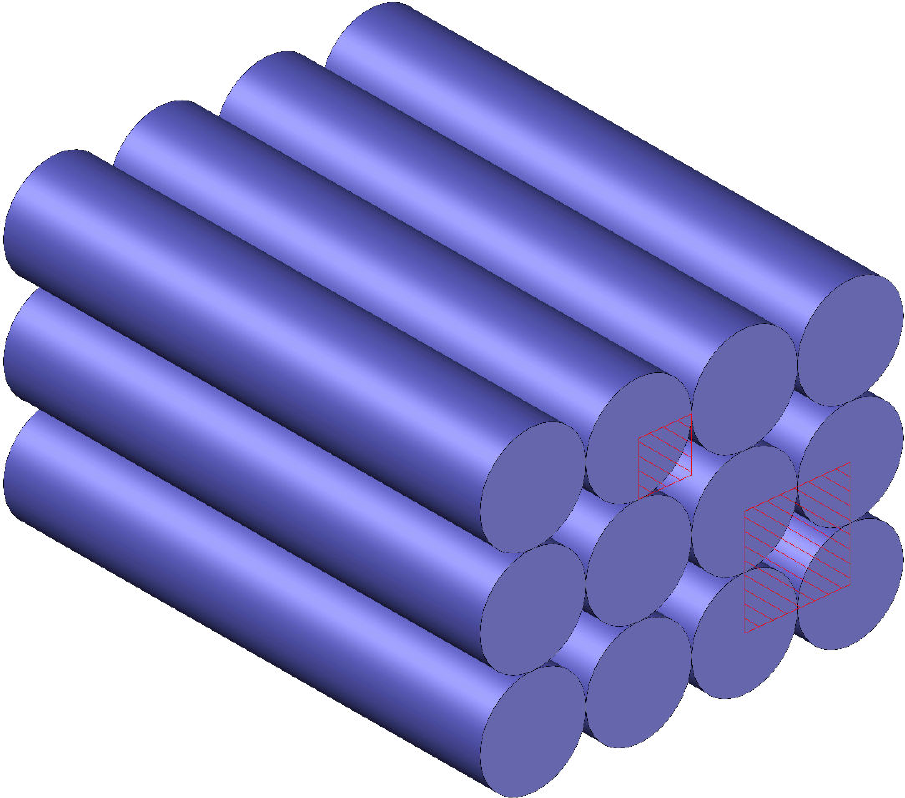}
}
\qquad
\subfloat[\emph{2D-Hexagonal packing, with the unit cell and the primitive cell shown red.}]{\label{fig:impaccamento-b}
\includegraphics[width=0.40\columnwidth,keepaspectratio]{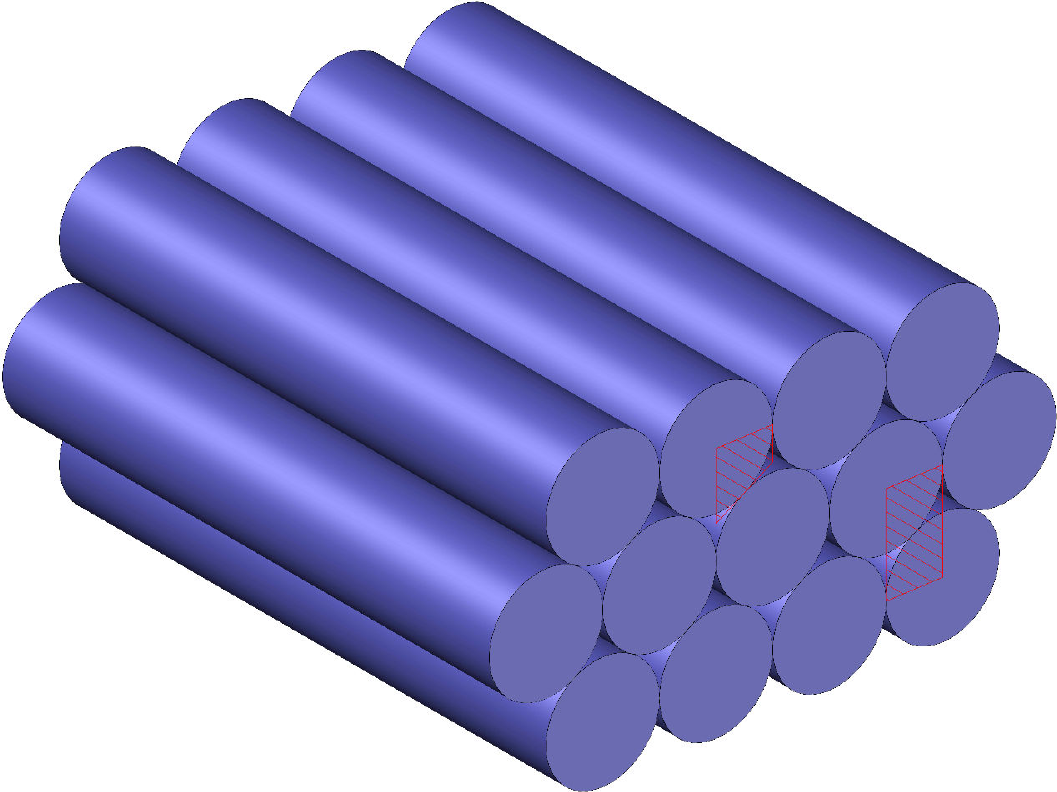}
}
\\
\subfloat[\emph{3D-cubic packing and its unit cell.}]{\label{fig:impaccamento-c}
\includegraphics[width=0.20\columnwidth,keepaspectratio]{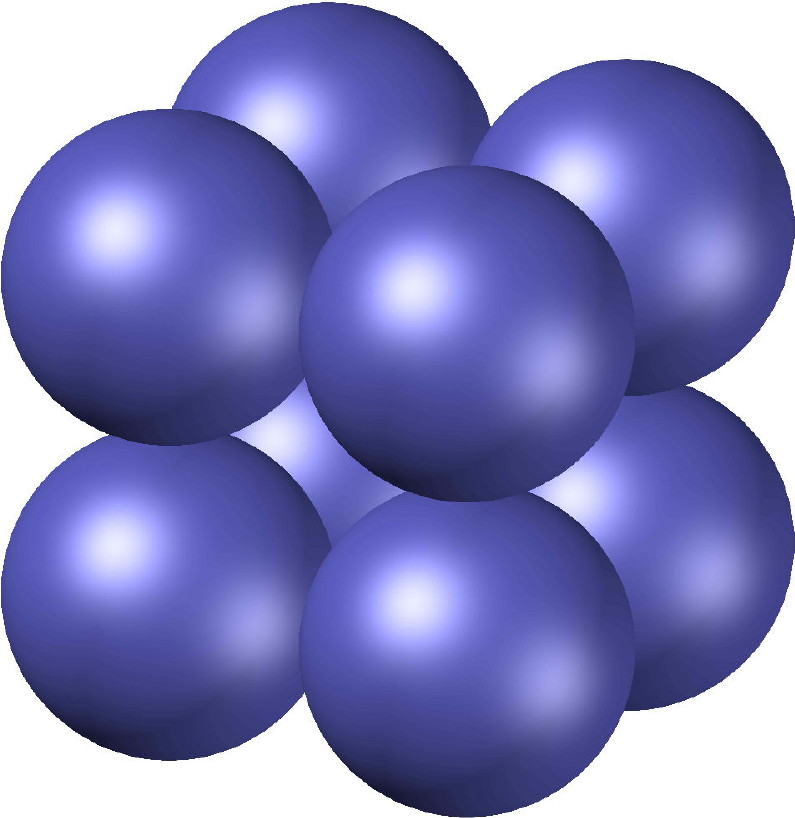}
\quad
\includegraphics[width=0.20\columnwidth,keepaspectratio]{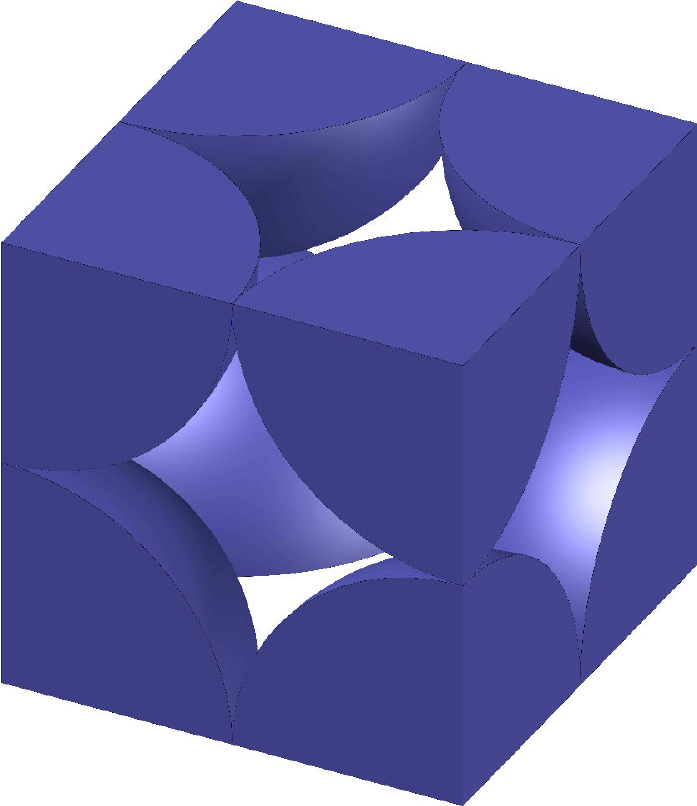}
}
\qquad
\subfloat[\emph{3D hexagonal packing and its unit cell.}]{\label{fig:impaccamento-d}
\includegraphics[width=0.20\columnwidth,keepaspectratio]{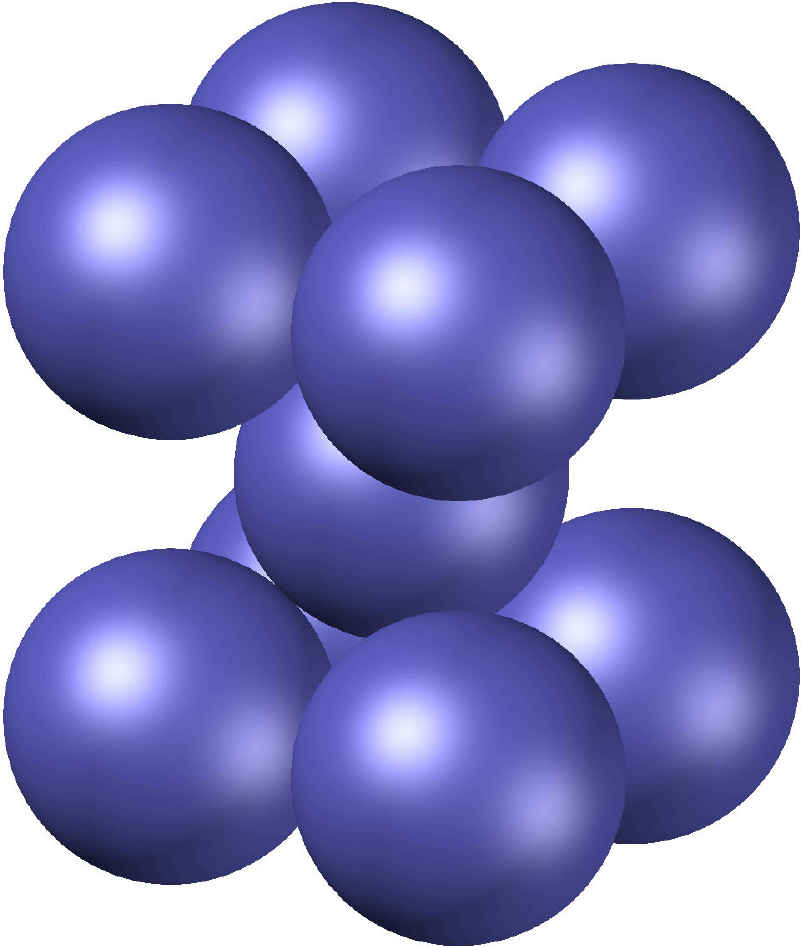}
\quad
\includegraphics[width=0.20\columnwidth,keepaspectratio]{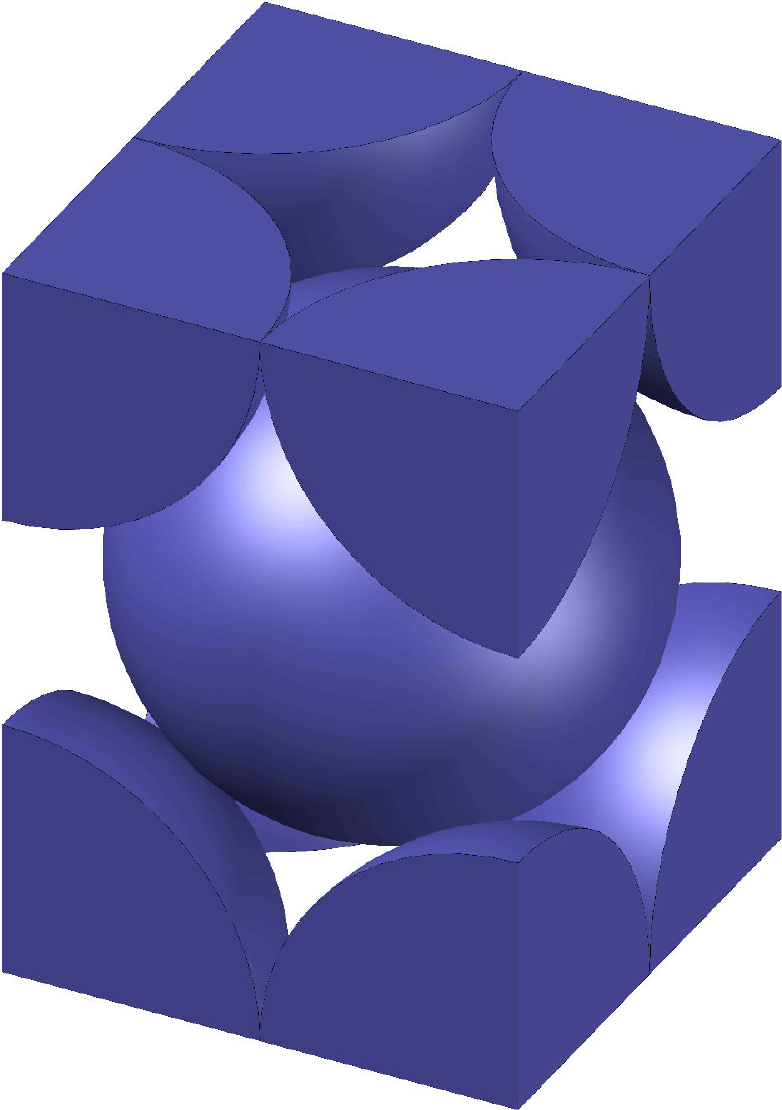}
}
\caption[Packing schemes]{
2D and 3D packing geometries idealizing the ceramic powder; the unit cells describing the periodicity and the primitive cells employed for the 2D numerical analysis are drawn red.
The primitive cells for the 3D analyses are shown in figure~\ref{fig:Ottavo-di-sfera-CAE}}
\label{fig:impaccamento}
\end{figure}
%} % Chiusura assegnazione nomi
%%%%%%%%%%--FINE-FIGURA--%%%%%%%%%%--FINE-FIGURA--%%%%%%%%%%%%%%%%%%%%%%%%%%%%%

%======================================================================================================================
% SEZIONE 3 : MODELLAZIONE MICROMECCANICA
%======================================================================================================================
\section[Micromechanical modelling]{Micromechanical modelling}
  \label{Sezione-03}

\begin{figure}[tp]
\centering
\subfloat[\emph{Model for the square packing.}]{\label{fig:quarto-a}
\includegraphics[width=0.349375\columnwidth,keepaspectratio]{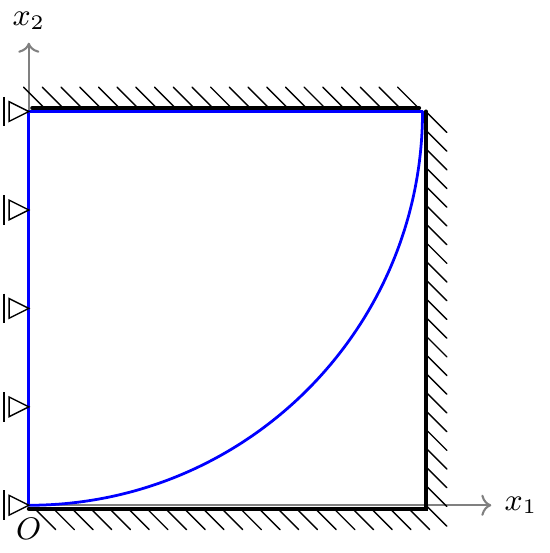}
}
\qquad
\subfloat[\emph{Model for the hexagonal packing.}]{\label{fig:quarto-b}
\includegraphics[width=0.349375\columnwidth,keepaspectratio]{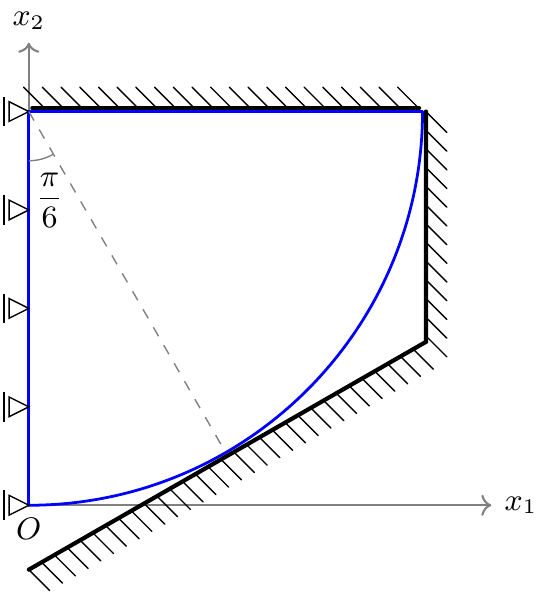}
}
\caption[Geometry and boundary conditions of the quarter of cylinder]{
The primitive cells employed for numerical simulations and depicted in figures~\ref{fig:impaccamento-a} and~\ref{fig:impaccamento-b} respectively are shown with the appropriate boundary conditions for the two-dimensional, plane strain packing schemes figures~\ref{fig:quarto-a} and~\ref{fig:quarto-b}.
}
\label{fig:Quarto-di-cilindro-CAE}
\end{figure}

\begin{figure}[tp]
\centering
\subfloat[\emph{Model for the cubic packing.}]{\label{fig:ottavo-a}
\includegraphics[width=0.27\columnwidth,keepaspectratio]{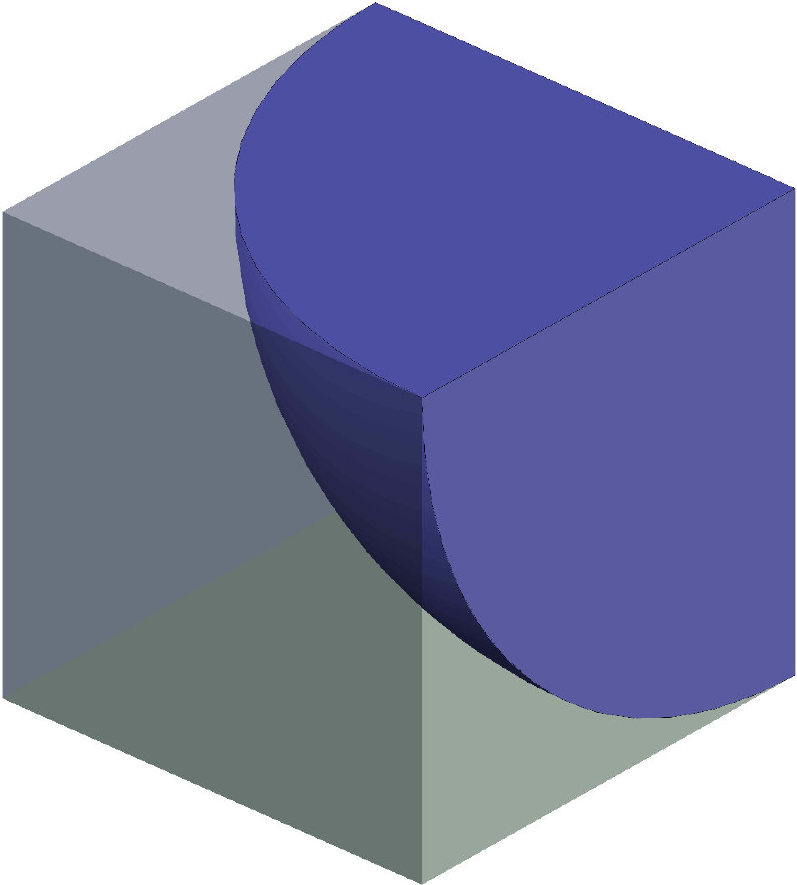}
}
\qquad
\subfloat[\emph{Model for the 3D hexagonal packing.}]{\label{fig:ottavo-b}
\includegraphics[width=0.27\columnwidth,keepaspectratio]{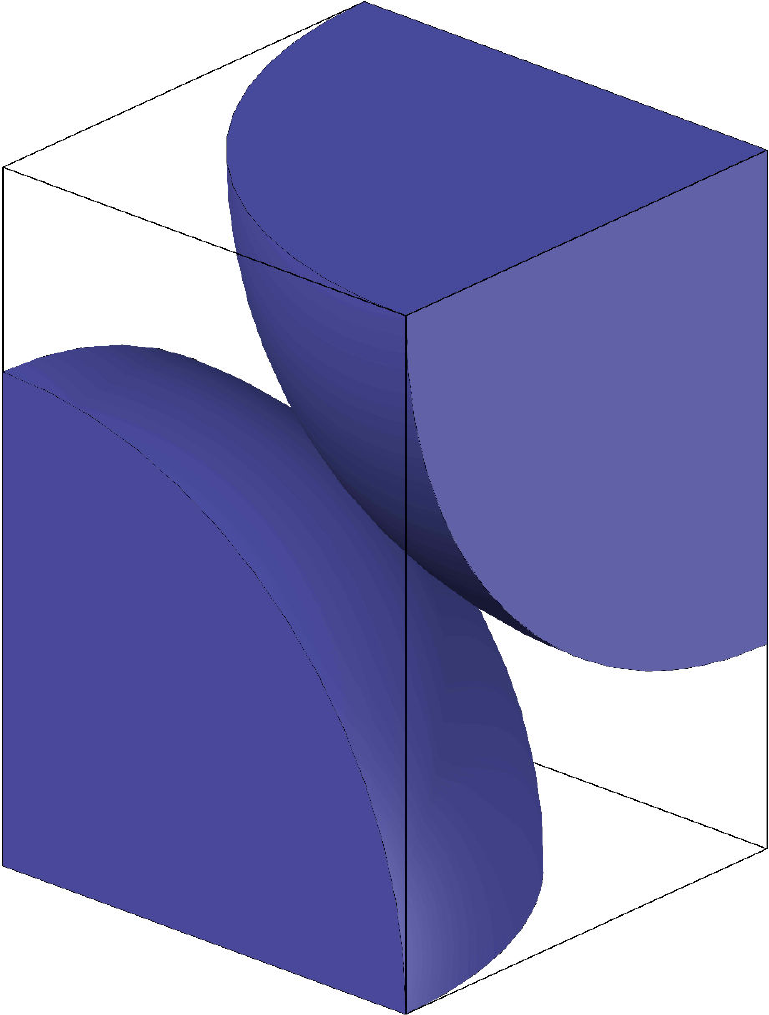}
}
\caption[Geometry of the eight of sphere]{The primitive cells employed for numerical simulations and depicted in figures~\ref{fig:impaccamento-c} and~\ref{fig:impaccamento-d} respectively are shown with the appropriate boundary conditions for the three-dimensional packing schemes for square (a) and hexagonal (b) geometries.
}
\label{fig:Ottavo-di-sfera-CAE}
\end{figure}

Square/cubic and hexagonal two-dimensional (grains are idealized as cylinders) and three-dimensional (grains are idealized as spheres) granule  dispositions are considered as representative of ceramic powders, figure~\ref{fig:impaccamento}.
Although at a first glance these geometries may appear far from the reality, they are usually considered to represent correctly the overall behaviour of granulates~\cite{Akisanya-Cocks-Fleck:1994, Akisanya-Cocks:1995, Ogbonna-Fleck:1995, Mesarovic-Fleck:2000, Redanz-Fleck:2001, Sridhar-Fleck-Akisanya:2001, Larsson-Olsson:2015, Harthong-Jerier-Doremus-Imbault-Donze:2009}.
For the considered packagings, symmetry allows the reduction into the primitive cells and the unit cells shown in figure~\ref{fig:impaccamento}. For 2D (a quarter of a solid disk) and 3D (two eighths of a solid sphere) the reduction is shown in figure~\ref{fig:Quarto-di-cilindro-CAE} and~\ref{fig:Ottavo-di-sfera-CAE}, respectively.
The grains are in contact with smooth and rigid surfaces and all contacts between grains (and hence with the rigid surfaces) are assumed to be frictionless.

%%%%%%%%%%--FIGURA--%%%%%%%%%%--FIGURA--%%%%%%%%%%%%%%%%%%%%%%%%%%%%%%%%%%%%%%%
%{\tikzset{external/figure name={Schema-quarti-di-cilindro_}} % Nome per le figure esternalizzate
%} % Chiusura assegnazione nomi
%%%%%%%%%%--FINE-FIGURA--%%%%%%%%%%--FINE-FIGURA--%%%%%%%%%%%%%%%%%%%%%%%%%%%%%

%%%%%%%%%%--FIGURA--%%%%%%%%%%--FIGURA--%%%%%%%%%%%%%%%%%%%%%%%%%%%%%%%%%%%%%%%
%{\tikzset{external/figure name={Schema-ottavi-di-sfera_}} % Nome per le figure esternalizzate
%} % Chiusura assegnazione nomi
%%%%%%%%%%--FINE-FIGURA--%%%%%%%%%%--FINE-FIGURA--%%%%%%%%%%%%%%%%%%%%%%%%%%%%%

\begin{figure}[tp]
\centering
\subfloat[]{\label{fig:legame-elastoplastico-a}
\includegraphics[width=0.25\columnwidth,keepaspectratio]{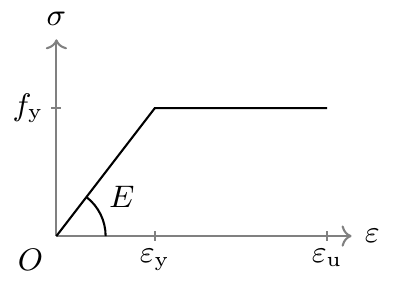}
}
\qquad
\subfloat[]{\label{fig:legame-elastoplastico-b}
\includegraphics[width=0.25\columnwidth,keepaspectratio]{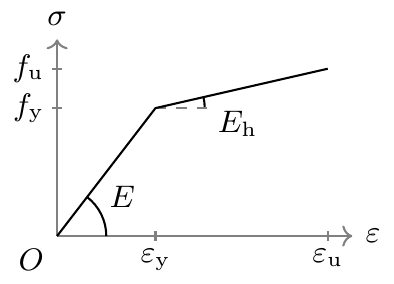}
}
\caption[Constitutive laws]{
Uniaxial stress/strain relation for ideal plastic behaviour~(\ref{fig:legame-elastoplastico-a}) and linear hardening~(\ref{fig:legame-elastoplastico-b}).
Both these constitutive equations are used in the modelling of the constitutive response of the ceramic grains.}
\label{fig:legame-elastoplastico}
\end{figure}

Reference is made to the ready-to-press commercial grade, 96\% pure, alumina powder (392 Martoxid KMS-96), one of the three investigated in Part~I of this study.
This powder has particles of~\SI{170}{\micro\metre} mean diameter, obtained through spray-drying, and possesses a high plastic formability, because particles are made up of an aggregate of microcrystals with a polymeric binder.
It is not known which constitutive equation models the material behaviour of the grains, except that it is an elastoplastic constitutive law.
For this reason, the simplest constitutive framework of plasticity is selected, namely, the von Mises yield surface with isotropic elastic part and perfectly plastic or isotropic hardening plastic law.
These models are characterised by an elastic Young modulus, a Poisson's ratio, a yield stress and a plastic hardening (which is null in the case of perfectly plastic behaviour), figure~\ref{fig:legame-elastoplastico}.
These parameters are unknown at the level of the single grain of powder.
Therefore, we have used the constitutive parameters as free parameters (although constrained to range within \lq reasonable' values) to find the best match with experiments.
For this reason it is anticipated that the constitutive parameters have been selected to be different for the different geometries.

%%%%%%%%%%--FIGURA--%%%%%%%%%%--FIGURA--%%%%%%%%%%%%%%%%%%%%%%%%%%%%%%%%%%%%%%%
%{\tikzset{external/figure name={Schema-legame-costitutivo_}} % Nome per le figure esternalizzate
%} % Chiusura assegnazione nomi
%%%%%%%%%%--FINE-FIGURA--%%%%%%%%%%--FINE-FIGURA--%%%%%%%%%%%%%%%%%%%%%%%%%%%%%

Loading under uniaxial \textit{strain} into a mould has been simulated with subsequent complete unloading and reloading (within the elastic range) under uniaxial \textit{stress} to determine the elastic Young modulus ad Poisson's ratio.
To this purpose, the horizontal plane bounding the upper part of the cell in figures~\ref{fig:Quarto-di-cilindro-CAE} and~\ref{fig:Ottavo-di-sfera-CAE} is assumed rigid and prescribed a vertical displacement, corresponding to a mean strain in the homogenized material.
After a certain mean strain (and corresponding relative density) has been reached, a total unloading is prescribed and the elastic Young modulus and Poisson's ratio are evaluated through an elastic unconfined reloading (corresponding to uniaxial average stress).

The numerical computations have been performed by means of the Abaqus Unified FEA$^\circledR$ software, in which the geometry described in figures~\ref{fig:Quarto-di-cilindro-CAE} and~\ref{fig:Ottavo-di-sfera-CAE} have been employed.
In particular, figures~\ref{fig:Quarto-di-cilindro-CAE} and~\ref{fig:Ottavo-di-sfera-CAE} are referring to the initial conditions of the confined uniaxial compaction of the cylindrical ceramic powder specimens (as can be deduced by the presence of frictionless and rigid side walls providing the confinement).
The extraction of the specimen from the mould (preceding the final unconfined uniaxial loading) is modelled through removal of the confining side walls.

In the two-dimensional models, structured meshes with CPE6H and CPE8H elements (6-node triangular and 8-node quadrilateral hybrid elements, respectively) were employed, thus yielding a total amount of \num{1894} elements and \num{5545} nodes for the quarter of cylinder, while in the three-dimensional models a free mesh with C3D4 elements (4-nodes tetrahedral linear element) was employed, thus yielding a total amount of \num{89252} elements and \num{18177} nodes for each eighth of the sphere.

The  confinement side walls were modelled as rigid elements and the contact interaction with the grain was modelled as frictionless contact in a large displacement regime between master (rigid surfaces) and slave (grain surface) surfaces, in which pressure-overclosure hard contact was used, allowing also for the separation of the bodies after contact.

The unconfined re-loading was prescribed through extremely small load increments, in order to accurately define the first part of the displacement-force curve from which the elastic Young modulus and Poisson's ratio of the unit cell can be deduced.

The material properties of the grains were selected differently for 2D and 3D, and for each packing scheme (as reported in table~\ref{tab:parametri-ottimizzazione-singola}).
These different selections were introduced to obtain the best fit between the elastic modulus of the unit cell and the experimental data (reported in Part~I of this study).

%======================================================================================================================
% SEZIONE 4 : EVOLUZIONE DELLA RIGIDEZZA
%======================================================================================================================
\section[The evolution of elastic stiffness from micromechanics]{The evolution of elastic stiffness from micromechanics}
  \label{Sezione-04}

\begin{figure}[tp]
\centering
\includegraphics[width=0.95\columnwidth,keepaspectratio]{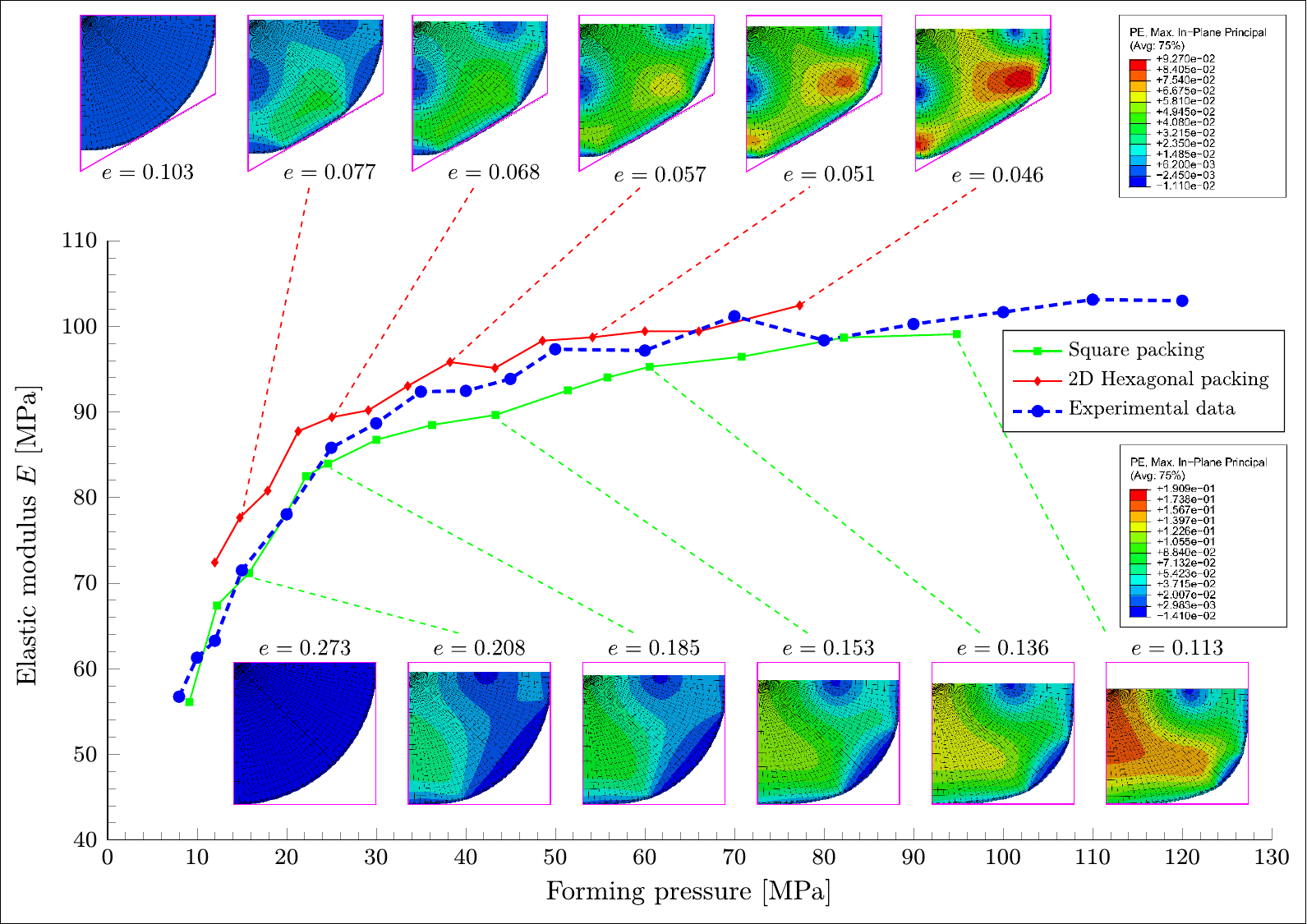}
\caption[Evolution of the elastic modulus of the two-dimensional cells]{
Simulated evolution of the elastic Young modulus with the forming pressure for a ceramic powder idealized as two-dimensional square and hexagonal packings of elastoplastic cylinders.
The material properties of the grains have been selected differently (table~\ref{tab:parametri-ottimizzazione-singola}) for the different packing schemes to obtain a best fit of the experimental results.
The shapes of the plastically deformed grains are reported at different stages of forming.}
\label{fig:Curve-Modulo-Elastico-Pressione-2D-Finale}
\end{figure}

\begin{figure}[tp]
\centering
\includegraphics[width=0.95\columnwidth,keepaspectratio]{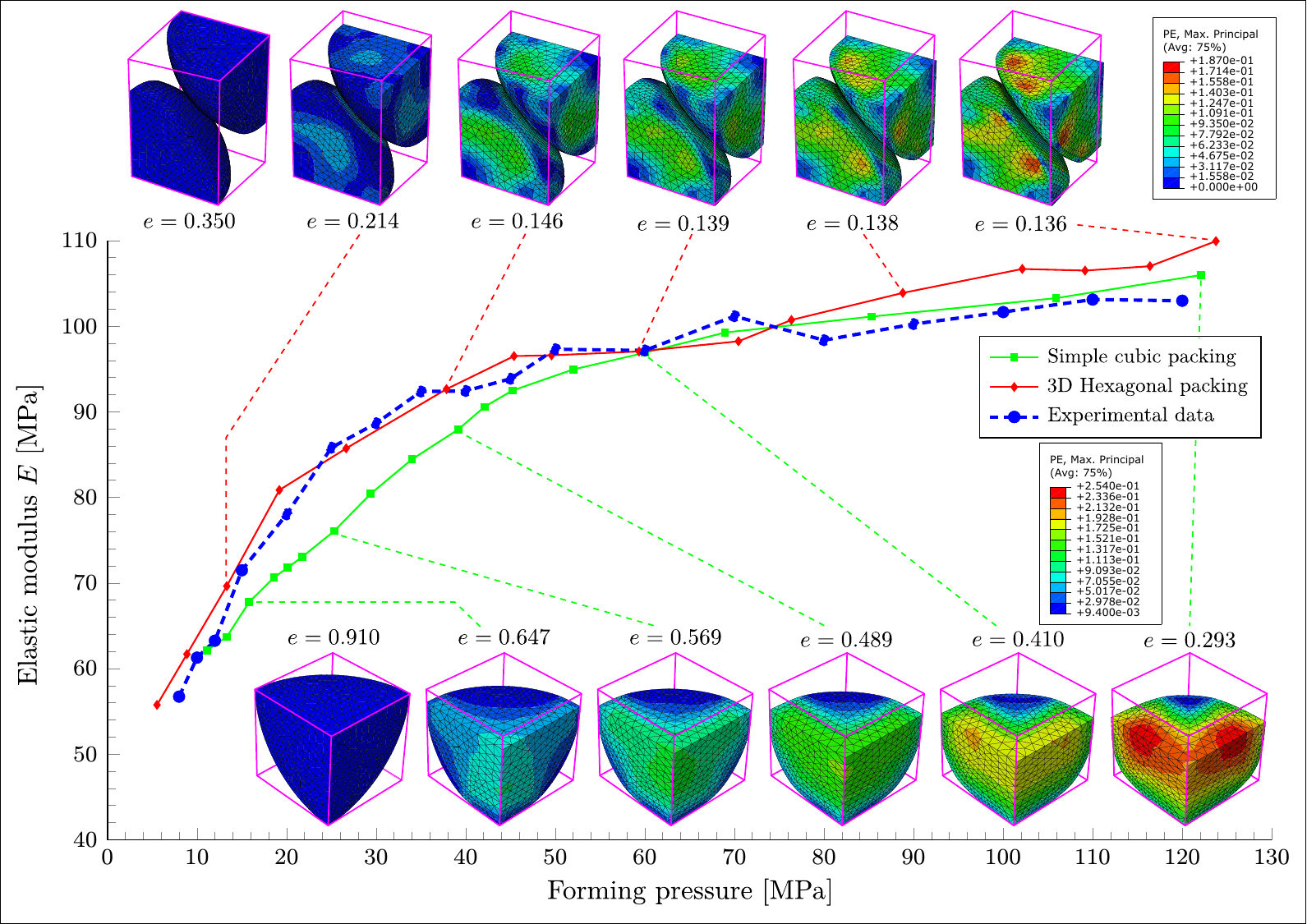}
\caption[Evolution of the elastic modulus of the three-dimensional cells]{
Simulated evolution of the elastic Young modulus with the forming pressure for a ceramic powder idealized as three-dimensional cubic and hexagonal packings of elastoplastic spheres.
The material properties of the grains have been selected differently (table~\ref{tab:parametri-ottimizzazione-singola}) for the different packing schemes to obtain a best fit of the experimental results.
The shapes of the plastically deformed grains are reported at different stages of forming.
}
\label{fig:Curve-Modulo-Elastico-Pressione-3D}
\end{figure}

The evolution of the elastic Young modulus $E$ with the forming pressure is reported in figure~\ref{fig:Curve-Modulo-Elastico-Pressione-2D-Finale} for the 2D simulations and in figure~\ref{fig:Curve-Modulo-Elastico-Pressione-3D} for the 3D cases.

%%%%%%%%%%--TABELLA--%%%%%%%%%%--TABELLA--%%%%%%%%%%%%%%%%%%%%%%%%%%%%%%%%%%%%%
\begin{table}[tp]
\centering
\caption[Summary of the material parameters]{Material parameters for the ceramic grains employed in the numerical simulations providing the best fit to the experiments}
\label{tab:parametri-ottimizzazione-singola}
\begin{tabular}{ccccc}
\toprule
\multirow{2}*{Grain mechanical properties} & \multicolumn{4}{c}{Packing geometry} \\
\cmidrule(lr){2-5}
                                     & Square                 & 2D Hexagonal            & Simple cubic           & 3D Hexagonal \\
\midrule
Elastic modulus $ E $                & \SI{400}{\mega\pascal} & \SI{350}{\mega\pascal}  & \SI{450}{\mega\pascal} & \SI{420}{\mega\pascal} \\
Poisson's ratio $ \nu $              & \num{0.35}             & \num{0.35}              & \num{0.35}             & \num{0.35}         \\
Yield stress $ f_{\textup{y}} $      & \SI{5}{\mega\pascal}   & \SI{6.56}{\mega\pascal} & \SI{4.5}{\mega\pascal} & \SI{4.2}{\mega\pascal} \\
Hardening modulus $ E_{\textup{h}} $ & \SI{320}{\mega\pascal} & \SI{335}{\mega\pascal}  & \SI{250}{\mega\pascal} & \SI{220}{\mega\pascal} \\
\bottomrule
\end{tabular}
\end{table}
%%%%%%%%%%--FINE-TABELLA--%%%%%%%%%%--FINE-TABELLA--%%%%%%%%%%%%%%%%%%%%%%%%%%%

%*************** Figure 2D ***************

%%%%%%%%%%--FIGURA--%%%%%%%%%%--FIGURA--%%%%%%%%%%%%%%%%%%%%%%%%%%%%%%%%%%%%%%%
%{\tikzset{external/figure name={Curve-Modulo-Elastico-Pressione-2D_}} % Nome per le figure esternalizzate
%} % Chiusura assegnazione nomi
%%%%%%%%%%--FINE-FIGURA--%%%%%%%%%%--FINE-FIGURA--%%%%%%%%%%%%%%%%%%%%%%%%%%%%%

%*************** Figure 3D ***************

%%%%%%%%%%--FIGURA--%%%%%%%%%%--FIGURA--%%%%%%%%%%%%%%%%%%%%%%%%%%%%%%%%%%%%%%%
%{\tikzset{external/figure name={Curve-Modulo-Elastico-Pressione-3D_}} % Nome per le figure esternalizzate
%} % Chiusura assegnazione nomi
%%%%%%%%%%--FINE-FIGURA--%%%%%%%%%%--FINE-FIGURA--%%%%%%%%%%%%%%%%%%%%%%%%%%%%%

In the figures the plastically deformed shapes of the grains are reported upon unloading and before the reloading imposed to determine the elastic Young modulus.

In both cases the material parameters listed in table~\ref{tab:parametri-ottimizzazione-singola} provide a close fit to experimental results (on alumina powder, see Part~I of this study) and highlight the elastoplastic coupling effect, which is given now a micromechanical basis.
Note that the differences in the geometries of the disposition of spheres and cylinders are \lq compensated' by the constitutive parameters of the grains, so that all the schemes provide a good representation of the variation of the elastic Young modulus with the forming pressure.
It is obvious that all geometries considered are idealization of a more complex reality; an improvement in the model would be to obtain direct information on the constitutive laws governing the mechanical behaviour of the grains.
This could be achieved with measures at the microscale, for instance nanoindentation, which are for the moment not available.

\begin{figure}[tp]
\centering
\includegraphics[width=0.683125\columnwidth,keepaspectratio]{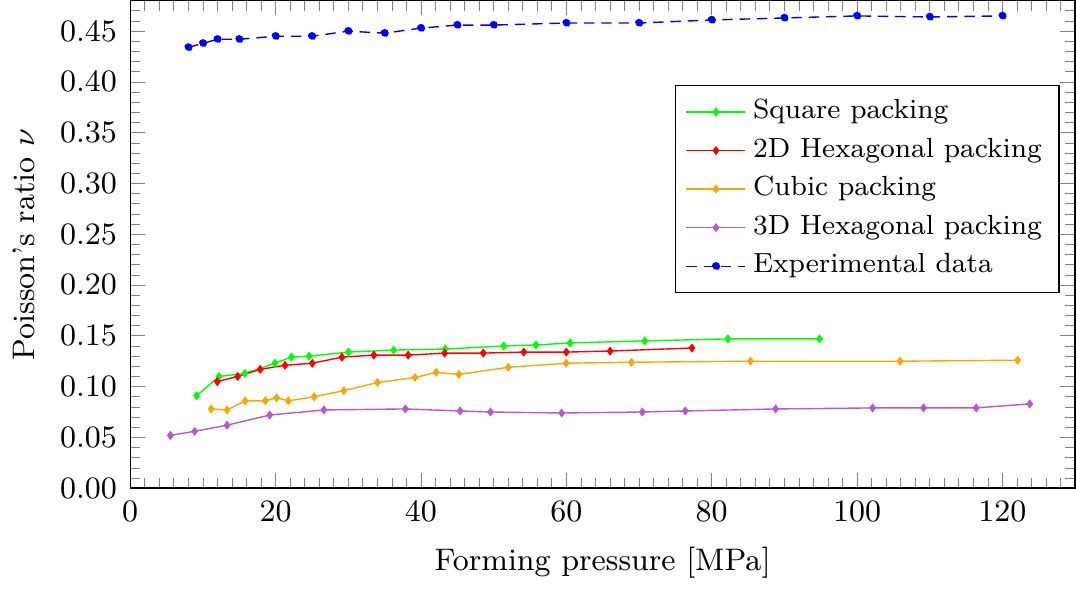}
\caption[Simulated Poisson's ratio as a function of the forming pressure during compaction]{
Simulated Poisson's ratio as a function of the forming pressure during compaction.
The ceramic powder is idealized as two-dimensional square and hexagonal packings of elastoplastic cylinders.
The material properties of the grains are the same used for the evaluation of the elastic modulus.
}
\label{fig:Curve-Coefficiente-Poisson-Pressione}
\end{figure}

The computation of the Poisson's ratio as a function of the forming pressure reported in figure~\ref{fig:Curve-Coefficiente-Poisson-Pressione} shows less agreement with theoretical results.
In fact, it can be noted that the qualitative behaviour is correct, thus predicting an increase in the Poisson's ratio with the forming pressure, but the computed values are definitely inferior to the experimental.
This is principally due to the fact that plastic strain reached in the numerical simulations is never as large as that reached in reality during powder compaction, so that a strong elastic release of deformation occurs at unloading in the simulations.

%%%%%%%%%%--FIGURA--%%%%%%%%%%--FIGURA--%%%%%%%%%%%%%%%%%%%%%%%%%%%%%%%%%%%%%%%
%{\tikzset{external/figure name={Curve-Coefficiente-Poisson-Pressione_}} % Nome per le figure esternalizzate
%} % Chiusura assegnazione nomi
%%%%%%%%%%--FINE-FIGURA--%%%%%%%%%%--FINE-FIGURA--%%%%%%%%%%%%%%%%%%%%%%%%%%%%%

\begin{figure}[tp]
\centering
\includegraphics[width=0.71875\columnwidth,keepaspectratio]{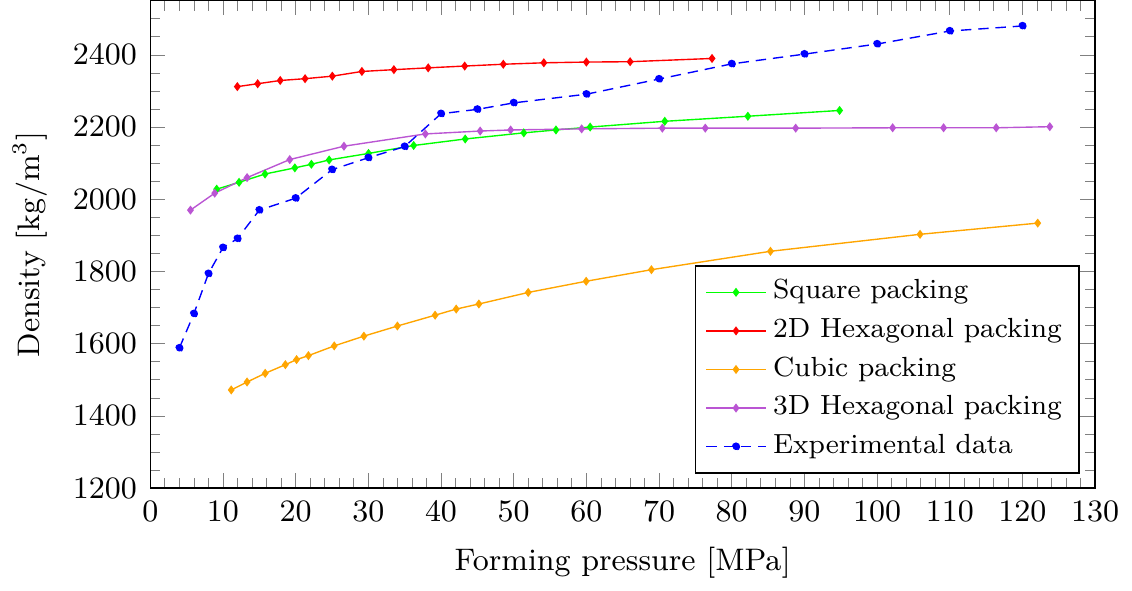}
\caption[Evolution of the unit cell density]{
Evolution of the unit cell density as a function of the forming pressure.
}
\label{fig:Curve-Pressione-Densita}
\end{figure}

Although providing practically the same elastic modulus/forming pressure relation, the different dispositions of grains yield different forming diagrams, in other words a different dependence of density of the greens on the forming pressure, as shown in figure~\ref{fig:Curve-Pressione-Densita}.
Coherently with results reported in figure~5 of Part~I of this study (upper part, C\&E fitting), the curves in  the figure has been obtained, by assuming a limit grain density $ \rho_{\infty}$ equal to~\SI{2500}{\kilogram/\metre^{3}}.

This figure shows that the best models are the square (two dimensional) and the 3D Hexagonal packings, while results from the 2D Hexagonal and cubic packings are less accurate.

%%%%%%%%%%--FIGURA--%%%%%%%%%%--FIGURA--%%%%%%%%%%%%%%%%%%%%%%%%%%%%%%%%%%%%%%%
%{\tikzset{external/figure name={Curve-Pressione-Densita_}} % Nome per le figure esternalizzate
%} % Chiusura assegnazione nomi
%%%%%%%%%%--FINE-FIGURA--%%%%%%%%%%--FINE-FIGURA--%%%%%%%%%%%%%%%%%%%%%%%%%%%%%

\begin{figure}[tp]
\centering
\includegraphics[width=0.68375\columnwidth,keepaspectratio]{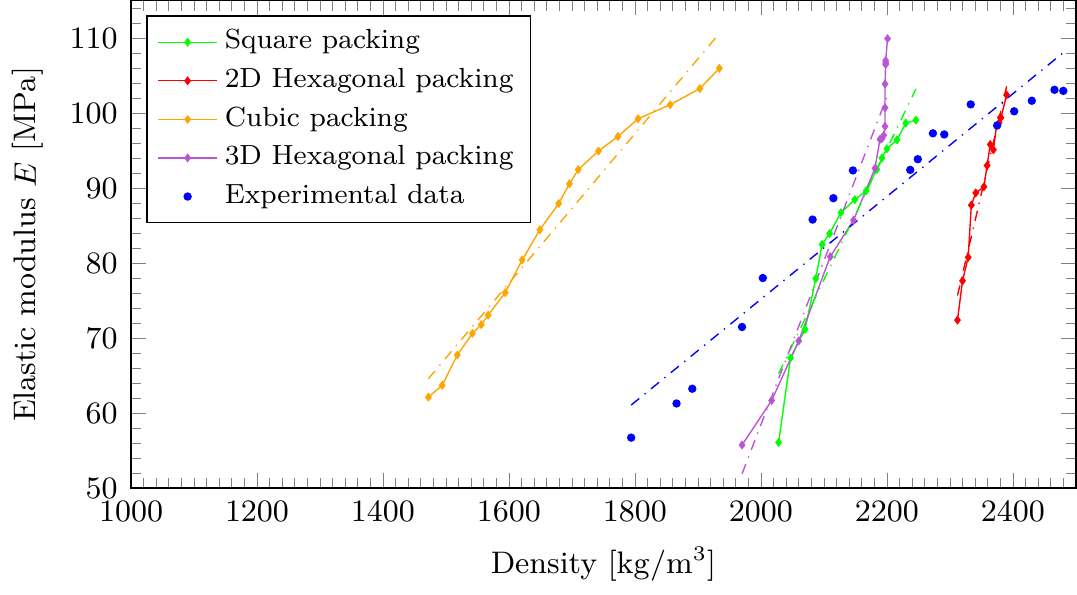}
\caption[Simulated values of elastic modulus as functions of the density during compaction]{
Simulated evolution of the elastic Young modulus with the density for a ceramic powder idealized as three-dimensional cubic and hexagonal packings of elastoplastic spheres.
The material properties of the grains have been selected differently (table~\ref{tab:parametri-ottimizzazione-singola}) for the different packing schemes to obtain a best fit of the experimental results.
}
\label{fig:Curve-Modulo-Elastico-Densita_0}
\end{figure}

The same conclusion can be drawn from figure~\ref{fig:Curve-Modulo-Elastico-Densita_0}, where the simulated evolution of the elastic Young modulus $E$ with the density is reported.
Here the micromechanical modelling shows a qualitative agreement and confirms the linear dependence found in Part~I of this study.

%*************** Figure proprietà elastiche - indice dei vuoti ***************

%%%%%%%%%%--FIGURA--%%%%%%%%%%--FIGURA--%%%%%%%%%%%%%%%%%%%%%%%%%%%%%%%%%%%%%%%
%{\tikzset{external/figure name={Curve-Modulo-Elastico-Densita_}} % Nome per le figure esternalizzate
%} % Chiusura assegnazione nomi
%%%%%%%%%%--FINE-FIGURA--%%%%%%%%%%--FINE-FIGURA--%%%%%%%%%%%%%%%%%%%%%%%%%%%%%

\begin{figure}[tp]
\centering
\includegraphics[width=0.683125\columnwidth,keepaspectratio]{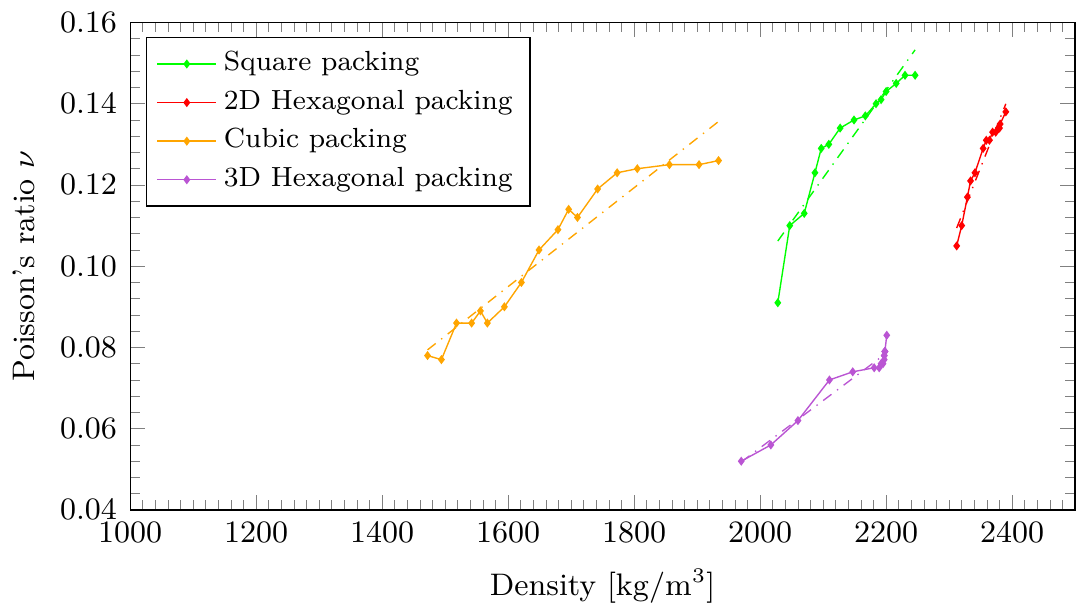}
\caption[Poisson's ratio as a function of the void ratio during compaction]{
Simulated evolution of the Poisson's ratio $\nu$ with the density for a ceramic powder idealized as three-dimensional cubic and hexagonal packings of elastoplastic spheres.
The material properties of the grains have been selected differently (table~\ref{tab:parametri-ottimizzazione-singola}) for the different packing schemes to obtain a best fit of the experimental results.
}
\label{fig:Curve-Coefficiente-Poisson-Densita_0}
\end{figure}

The evoulution of the Poisson's ratio with the density, reported in figure~\ref{fig:Curve-Coefficiente-Poisson-Densita_0} shows again only a qualitative agreement with the experimental data and a linear behaviour, but the discrepancy already visible in figure~\ref{fig:Curve-Coefficiente-Poisson-Pressione} is again found.

%%%%%%%%%%--FIGURA--%%%%%%%%%%--FIGURA--%%%%%%%%%%%%%%%%%%%%%%%%%%%%%%%%%%%%%%%
%{\tikzset{external/figure name={Curve-Coefficiente-Poisson-Densita_}} % Nome per le figure esternalizzate
%} % Chiusura assegnazione nomi
%%%%%%%%%%--FINE-FIGURA--%%%%%%%%%%--FINE-FIGURA--%%%%%%%%%%%%%%%%%%%%%%%%%%%%%

In conclusion, it can be pointed out that the micromechanical approach fully confirms the experimental finding that the elastic stiffness increases with the density of the green, even if there is only a partial quantitative agreement with the experimental data.

%======================================================================================================================
% SEZIONE 5 : CONCLUSIONI
%======================================================================================================================
\section[Conclusions]{Conclusions}
  \label{Sezione-05}

A micromechanical approach has been developed to explain the increase of elastic stiffness with the density of the greens observed during forming of ceramic powders (experiments have been reported in Part~I of this study).
Although the mechanical characteristics of the single granule are not known, a reasonable elastoplastic model for this has been employed, so that it has been possible to consider idealized 2D and 3D geometrical configurations of grains (represented as circular cylinders or spheres) and to load and unload a representative unit cell. In this way, the variation with forming pressure was determined of  elastic modulus, Poisson's ratio, and density.

It has been shown that the micromechanical analysis can explain both qualitatively and quantitatively the increase of the Young modulus with forming pressure, while the behaviour of the Poisson's ratio is only qualitatively confirmed. Moreover, a validation is provided for the linear dependence of the two elastic parameters on the relative density of the material (found in Part~I of this study).

%======================================================================================================================
% SEZIONE 6 : RINGRAZIAMENTI
%======================================================================================================================
\section*{Acknowledgments}
The authors gratefully acknowledge financial support from  the European FP7 programs INTER\-CER-2 PIAP-GA-2011-286110-INTERCER2 (L.P.A., D.C., and D.M.), ERC-2013-ADG-340561-INSTABILITIES (D.B.), and FP7-PEOPLE-2013-ITN-PITN-GA-2013-606878-CERMAT2 (A.P.).

%======================================================================================================================
% BIBLIOGRAFIA
%======================================================================================================================
% \phantomsection                  % Attivare questa opzione solo se è caricato il pacchetto hyperref
% \printbibliography               % Per stampare tutta la bibliografia
\bibliographystyle{unsrt}
\bibliography
{Bibliografia}

\end{document}